\documentclass[12pt, letterpaper]{article}
\usepackage[utf8]{inputenc}
\usepackage{amsmath}
\usepackage{graphicx}
\graphicspath{ {./images/} }
\usepackage[utf8]{inputenc}
\usepackage{amssymb}
\usepackage{changepage}
\begin{document}

%\title{Search for the  \text{H} \rightarrow $\text{W}^{+} %\textbf{W}^{-}$ process at the LHeC Experiment}

%\author[mymainaddress]{Hollis Williams}

\title{Search for the  $\text{H} \rightarrow$ W$^{+}$ W$^{-}$ Process at the LHeC Experiment}
\author{Hollis Williams}
\maketitle
%\title{
%Search for the  \text{H} \rightarrow $\text{W}^{+}$ W$^{-}$ %process at the LHeC Experiment}

\begin{abstract}
We consider the decay of the Higgs boson to W$^{+}$ W$^{-}$ at a proposed Large Hadron Electron Collider and determine the likelihood of detecting a signal for the Higgs mass from its decay product W jets by imposing cuts to select candidate jet pairs and optimizing the value of the angular separation $\Delta R$.  It was found that at the LHeC experiment (CM energy $\sqrt{s}=1.3$ TeV and luminosity of 100 fb$^{-1}$ per year), the highest efficiency is obtained with $\Delta R = 0.4$, along with a selection scheme of $|\Delta \eta| <1, 10<m<85$ GeV, $p_T$ of jets 1 and 2 between $10 - 20$ GeV and $p_T$ of jets 3 and 4 $>10$ GeV: this led to an efficiency between $7.1 - 7.5 \%$ for finding the invariant 4-jet mass in a mass region $<140$ GeV.  Under signal-to-background comparison, the signal showed a $3.8 \sigma$ excess compared to the charged current W$^{-}$ background.

\end{abstract}

\section{Introduction}
\label{}
\noindent
A Higgs boson is an excitation of the Higgs field, the field from which fundamental fermions and massive gauge bosons acquire their masses (with the fermions getting their mass from Yukawa coupling to this field and the gauge boson obtaining it via the Higgs mechanism).  The major puzzle which the Higgs mechanism solves is the mass of the electroweak gauge bosons.  In order to experimentally study the Higgs in more detail and determine more of its properties, it would be necessary to produce a large number of them on a reliable basis.  The LHC is not suitable for this purpose as it has a large QCD background, making it difficult to separate the signal associated with a Higgs decay channel from other interactions which involve multi-jet final states.  Although the discovery of the Higgs boson was a key discovery of the LHC, we still lack detailed understanding of its properties and couplings (including whether it is really a fundamental scalar in the sense that leptons are fundamental, or whether it is a composite particle).  The branching ratios also need to be measured rigorously and checked against the predictions of the SM and it remains to be seen whether the measured total Higgs decay cross section can be accounted for using only the particles of the SM. 

By studying how it couples to its decay products we may also uncover properties which are unexpected or not explained in the SM.  It is known that the Higgs has even parity and zero spin and hence it represents an (apparently) fundamental scalar, unlike scalar mesons, which are hadronic composites.  Its mass is a free parameter of the SM given by $m_{H}=2\lambda v^2$.  It is a neutral particle and as a consequence of its role in generating mass, it couples to mass.  In theory, the Higgs can decay to any other particle in the SM but the coupling is proportional to mass, so the largest branching ratio should be to the most massive particle which is kinematically accessible.  It follows that the decay mode with the largest branching ratio is $b\overline{b}$.  In this work, we study the possiblity of using an ep collider to search for Higgs boson production via Higgs decay to W boson pairs.   

At the LHeC, the most probable mechanism for Higgs boson production is a charged current or netural current interaction via W or Z boson fusion, resulting in a Higgs boson, a jet and an electron neutrino [2].

\begin{figure}
\centering
\includegraphics[width=11cm, height=5cm]{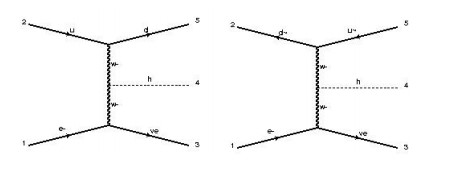}
\newline
Figure 1: Tree Level Feynman Diagrams contributing to Higgs Boson Production via W Boson Fusion
\end{figure}

\noindent
In general, identification of the Higgs at the LHC via hadronic decay of the W boson is not considered viable because it is difficult to distinguish final state jet decays of the Higgs from the huge number of other events at the LHC which involve multi-jet final states.  At the LHeC the QCD background is cleaner by a factor of 100 and the DIS final state is clean.  It is possible to detect the W boson through its leptonic decays (such as $e^{+}\nu$) but these have relatively small branching ratios.  The branching ratio for the W boson decay to two jets is dominant with the ratio of $W^{+} \rightarrow q \bar{q}$ being $(67.41 \pm 0.27)\:\%$ [8].  Given the abundance of this branching fraction, there is the potential to not only study $H \rightarrow \text{WW}^*$ in more detail at the LHeC but also to use it to measure Higgs boson production and so create a precision Higgs factory where the Higgs can be created on a reliable basis.  There is interest in studying the HWW coupling because, unlike at the LHC where the HZZ and HWW couplings cannot be separated, the two are separate at the LHeC and the latter could have contributions which are not explained by the SM.  A more detailed knowledge of Higgs couplings is also necessary to determine if the fundamental fermions really do obtain their masses via Yukawa coupling to the Higgs field.

Studying the $H \rightarrow W^{+} W^{-}$ process in detail relies on the possibility of tagging on W jets (as has already been done at LHC).  The production of a W boson and another virtual W boson is followed by hadronic decays of both Ws, resulting in four or five W jets in the final state (we will leave aside the leptonic decays). When the W boson is very energetic, the two jets will be close together and will merge, meaning that we actually end up reconstructing one single jet characterised by a two-prong structure. The analysis of this channel relies on the possibility of being able to distinguish the W jets from jets due to quarks and gluons produced in strong interactions [5].

\section{Large Hadron Electron Collider}

\noindent
Physicists are currently exploring options for a next-generation collider at the energy frontier: two possibilities are a new electron-positron collider or a LHeC (Large Hadron-Electron Collider).  The former would be similar to the LEP (Large Electron-Positron Collider) and the latter would be similar to HERA at DESY but with a greater centre of mass energy compared to CM energy of 318 GeV at HERA. The advantage of an ep collider is that it offers the opportunity to observe phenomena which would be observed in a pp collider with a cleaner decay environment and reduced contamination from unwanted multi-jet final states.  One important aspect of the LHeC which separates it from the LEP is that it complements the LHC: the LHeC would provide an electron beam between 60 and 140 GeV (compared to 27.5 GeV for the lepton beam at HERA) which would be collided with the intense hadron beams already provided by the LHC.  This would increase the kinematic range by a factor of twenty for $Q^2$ and inverse $x$ and there would be an increase over the integrated luminosity of HERA of two orders of magnitude with a luminosity of $10^{33}$ $\text{cm}^{-2}\:\text{s}^{-1}$.  The LHeC could potentially be realized as a ring-ring or a ring-linac configuration.  In the ring-ring configuration the same geometry is used for both components and the technology of the ring setup has been extensively studied at HERA and LEP.  The electrons are accelerated in a ring whereas in the ring-linac configuration the electrons are accelerated to the required energies in a linear accelerator before being collided with the protons travelling around the LHC.  The process of generating intense lepton beams in a storage ring is well-understood. However, the linac-ring configuration has the advantage that the infrastructure of the linac only meets with the ring in the vicinity of the interaction vertex, minimising interference due to hadron beams [5].  An initial 500 MeV electron bunch originating at the injector is accelerated to 10 GeV in each linac, leading to a final energy of 60 GeV at the interaction vertex after passing through the entire setup three times.  The 60 GeV beam is then collided with the proton beam from the LHC.

\section{Simulated Samples}
\noindent
The search was performed using a simulated LHeC experiment with  $\sqrt{s}=1.3$ TeV ep collisions and luminosity of 100 fb$^{-1}$ per year.  In almost all of the analysis of the jet kinematics there was a $p_T$ cut on the jets which would assist in jet reconstruction.  Low $p_T$ jets are typical when the jets are formed via hadronization of QCD radiation.  These background jets can be due to quarks or gluons emitted by particles inside the signal jet which then fragment and hadronize to form new jets [1].  

These false jets typically have smaller values of $p_T$ and so can be removed with a cut on transverse momentum of the jets.  When reconstructing the jets we must also consider the separation of the jets, their size and the algorithm used.  An algorithm which is in use at LHC (and in simulations of LHeC for consistency purposes) is the anti-$k_T$ algorithm along with a distance parameter.  A typical way of trying to categorize signal jets which emerge from a decay is to study how many of them have merged or separated configurations.  For example, we could consider a dijet with $\Delta R = 0.4$ as being separated, whereas a pair of jets with separation below this overlap and merge to form one jet.  $\Delta R$ is defined as follows:

\[\Delta R = \sqrt{(\Delta \eta)^2 + (\Delta \phi)^2}, \]

\noindent
where $\eta$ is the pseudorapidity and $\phi$ is the usual angular coordinate.  The anti-$k_T$ algorithm is an example of a sequential recombination jet algorithm for jet reconstruction.  These algorithms normally have as their parameter the power of the energy scale in the distance measure, and the 'anti' comes from the fact that the power is negative, as opposed to the ordinary $k_T$ algorithm.  

\[d_{ij}=\text{min}(k_{ti}^{2p},k_{tj}^{2p})\frac{\Delta_{ij}^2}{R^2}.\]

\noindent
Generation of events and cross sections (and numerical evaluation of  relevant matrix elements) was carried out via MadGraph5 which was used for generation of signal and background events at the LHeC [1].  The partons produced by MadGraph5 were assigned 4-vectors and were then showered by Pythia, an event generator which simulates the hadronisation and decay of the showers [7].  The resulting events were assigned 4-vectors and interfaced to Delphes, a detector package which includes simulations of systems to trigger on tracks and simulations of calorimeters and muon detectors.  Delphes analyses events generated by Pythia and creates a dataset as output which can be used for reconstruction [3]. 
    
\section{Event Selection}
\subsection{ Parton Level Plots}

\noindent
We begin by checking that our samples are for the WW* decay where one of the W bosons is off mass shell, and not WW, where both W bosons are on-shell.  The latter case makes the Higgs decay to two W bosons effectively impossible since $m_H < 2m_W$.  The partonic mass distributions were plotted for W+ and W$-$ and shown to be almost the same and in both cases a sharp peak was observed around the mass of the W boson (approximately 80 GeV) along with a smaller peak between 10 - 50 GeV, reaching a highest point between 30 - 40 GeV.  Since the off-shell W boson has a smaller mass this confirmed that the samples generated WW*.  It was also confirmed that a plot of the Higgs mass at parton level led to a peak at the Higgs mass as expected (around 125 GeV).
\newline

\includegraphics[width=14cm, height=8cm]{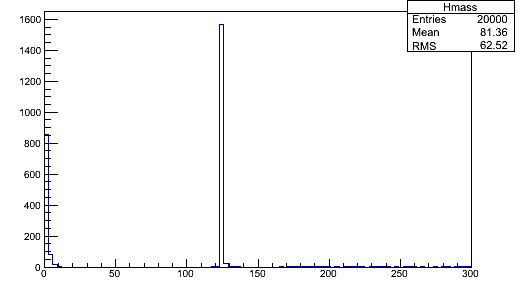}
\newline
Figure 2: Parton Level Plot of Higgs Mass

\noindent
\subsection{W-jets Selection}

\noindent
A selection criterion on the transverse momentum of the jets was required to remove a large number of the false jets which are present due to background QCD radiation.  To determine this criterion, the invariant mass of the two leading jets in each event (jets 1 and 2) was plotted.   Kinematically, it was expected that the W jets in an event could be detected via the jet pairs composed of the leading jets and so the jet pair composed of leading jets 1 and 2 should reconstruct to have the invariant mass of an on-shell W boson and the pair composed of leading jets 3 and 4 should reconstruct to the invariant mass of an off shell W boson.  A cut of $p_>30$ GeV was found to remove a large number of events with background jets whilst retaining the structure of the W peak, hence it was determined that this cut would be used as a maximum for removing background events, being lowered as necessary to detect the off shell W* (since a cut of $p_{T}>30$ GeV is rather high to see a signal from a W* between 10 and 50 GeV).  In fact, a cut of $p_{T}>30$ on all jets is very high for charged-current DIS and parton level observation showed that  the transverse momentum $p_{T}$ of the generated Higgs boson peaked at 50 GeV, wheras the $p_{T}$ of the on-shell W boson would be expected to peak at 40 GeV.
\newline
\subsection{Signal Selection}

\noindent
To reconstruct the Higgs mass, a cut of $|\Delta \eta|<1$ (where $\eta$ is the pseudorapidity) was imposed the for  difference between two jets in a jet pair.  The four $W$-tagged jets being used to reconstruct the Higgs mass can obviously appear in multiple combinations (for example, we could have the jet pairs (1,3) and (2,4) both in the necessary mass window, but not (1,2) and (3,4)).  Ideally one would like to consider the $\eta$ differences for all the jets in an event and not just the 4 leading jets, so the masses of all possible permuations of jet pairs in an event was considered where each permutation contains 4 jets.  Analysis confirmed that the signal for the Higgs boson could be improved by taking cuts of $p_{T} > 20$ GeV on jets 1 and 2, $p_{T} > 10$ GeV on jets 3 and 4, $|\Delta \eta|<1$ and invariant mass $m$ of jets 3 and 4 larger than 10 GeV.  The histogram shows a fairly high number of entries around the Higgs mass region with these selection criteria.

\includegraphics[width=14cm, height=8cm]{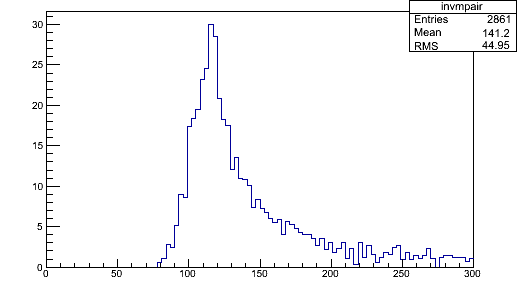}
\newline
Figure 3: Mass Distribution for the 4-Jet System (Cuts of $p_T > 10 $ on (3,4), $m>10 $)

\noindent
\linebreak
Up to this point, all samples used for the analysis assume a value of angular separation between jets of $\Delta R=0.4$.  This is likely optimal for the types of jet which are being studied.  This assumption was confirmed by repeating the analysis for other values of $\Delta R$ and finding that other values apart from $\Delta R = 0.4 $ or $\Delta R = 0.5$ reduce the number of events falling into the Higgs mass region.  The actual optimal value was confirmed later via selection efficiencies.  

\includegraphics[width=14cm, height=8cm]{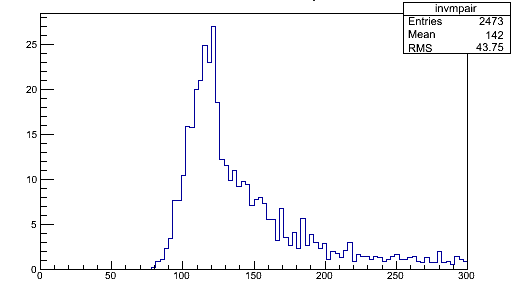}
\newline
Figure 4: Mass Distribution for the 4-Jet System ($\Delta R =0.5$)
\newline
\includegraphics[width=14cm, height=8cm]{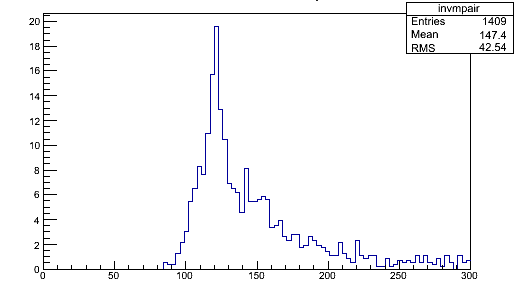}
\newline
Fig 5: Mass Distribution for the 4-Jet System ($\Delta R =0.7$)
\newline

\section{Signal to Background Comparison}
\noindent
A signal to background comparison was difficult for a study of this kind as the main background was due to multi-jet final states from charged-current DIS or from photoproduction of multi-jets.  Since it is non-trivial to produce high statistics of multi-jets, it would be normal in this situation to make cuts at parton level and then optimize the $\Delta R$ parameter for the separation between candidate W jets.  However, it was still desirable to have a preliminary background sample to show that it was suppressed by the proposed cuts on the signal sample (for example, a sample of background W${-}$ jets).  Such a smaple should also take account of the asymmetry in the WW$^{*}$ decay.  The analysis was repeated with the BG sample over the same number of events and a comparison made with the previous results. 
\newline
\includegraphics[width=16cm, height=5cm]{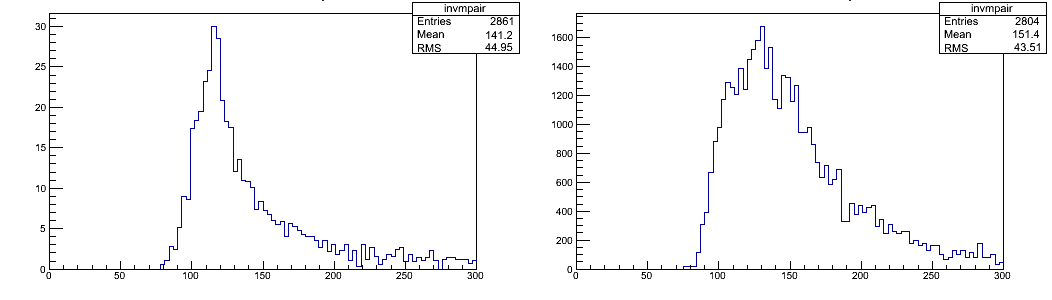}
\newline
Figure 6: Signal to Background Comparison
\newline
\noindent
\linebreak
Even accounting for normalization, the suppression of background events does not look promising so a stricter mass cut is required on the 4-jet of $m<130$ GeV or $m<140$.  
\newline
\newline
\includegraphics[width=16cm, height=5cm]{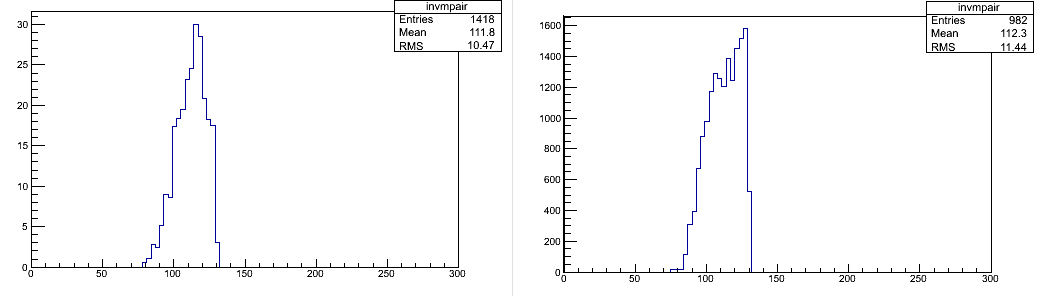}
\noindent
\newline
Figure 7: Signal to Background Comparison with Mass Cut $m<130$ GeV on 4-Jets
\newline
\linebreak
A strict cut is required on the mass of the 4-jet in order to see a good signal-to-background ratio: 1.44 for $m<130$ and 1.28 for $m<140$ compared to 1.02 for no mass cut.
\newline
\subsection{Selection Efficiencies}
\noindent
Efficiencies were calculated for various cuts on $p_T$ and $m$ at $\Delta R = 0.4$.  The selection efficiency was defined as the ratio between the number of events passing the selection criteria with 4-jet masses below 140 GeV and the total number of events.  $p_{Tij}$ denotes the $p_T$ cut on the jet pair composed of jets $i$ and $j$ and $m$ is the lower mass cut on the 4-jet.  
\newline
\newline
\begin{tabular}{ |p{6cm}|p{2cm}|p{3cm}|p{2cm}|
  }
\hline

\hline
Cuts& Signal &Background &S-B Ratio\\
\hline
$p_{T12}>20,p_{T34}>20,m>10$ & 186 &89& 2.09\\
$p_{T12}>20,p_{T34}>20,m>20$ & 161 &78 &2.06\\
$p_{T12}>20,p_{T34}>15,m>20$ & 472 &302 &1.56\\
$p_{T12}>20,p_{T34}>15,m>10$ & 594 &391 &1.52\\
$p_{T12}>20,p_{T34}>10,m>10$ & 1644 &1282& 1.28\\
$p_{T12}>20,p_{T34}>10,m>20$ & 1140 &845 &1.35\\
$p_{T12}>15,p_{T34}>10,m>20$ & 1192 &885 &1.35\\
$p_{T12}>15,p_{T34}>10,m>10$ & 1731 &1353 &1.28\\
$p_{T12}>10,p_{T34}>10,m>20$ & 1195 &887 &1.35\\
$p_{T12}>10,p_{T34}>10,m>10$ & 1738 &1363 &1.28\\
\hline
\end{tabular}
\newline
\newline
Table 1: Number of Events Passing Selection Criteria ($m<140$ GeV)
\newline
\linebreak
One obvious conclusion from the table is that lowering the mass cut from $m>20$ to $m>10$ increases the selection efficiency but lowers the signal-to-background ratio.  The difference in signal-to-background ratio in this case is negligible (between 0.03 and 0.07) and so the focus was placed on the candidate cut sets which produced the greatest number of events, especially since signal-to-background was not a major part of the study.  The cuts which produced the largest efficiencies were $p_{T12}>20,p_{T34}>10,m>10$, $p_{T12}>15,p_{T34}>10,m>10$ and $p_{T12}>10,p_{T34}>10,m>10$.  The analysis was run for the above optimal cut sets over samples with $\Delta R$ going from 0.3 to 0.7.  
\newline
\newline
\begin{tabular}{ |p{5cm}|p{1cm}|p{1cm}|p{1cm}|p{1cm}|p{1cm}|
  }
\hline

\hline
Cuts& 0.3 &0.4 &0.5 &0.6 &0.7\\
\hline
$p_{T12}>20,p_{T34}>10,m>10$ & 715 &1644& 1416 & 1027&720\\
$p_{T12}>15,p_{T34}>10,m>10$ & 757 &1731 &1488& 1098&765 \\
$p_{T12}>10,p_{T34}>10,m>10$ & 758 &1738 &1494& 1110&777\\
\hline
\end{tabular}
\newline
\newline
Table 2: Number of Events Passing Selection Criteria for $\Delta R = 0.3 - 0.7$
\newline
\linebreak
The analysis was repeated for a mass cut of $<130$ GeV on the 4-jets.
\newline
\newline
\begin{tabular}{ |p{6cm}|p{2cm}|p{3cm}|p{2cm}|
  }
\hline

\hline
Cuts& Signal &Background &S-B Ratio\\
\hline
$p_{T12}>20,p_{T34}>20,m>10$ & 132 &43& 3.07\\
$p_{T12}>20,p_{T34}>20,m>20$ & 113 &34 &3.32\\
$p_{T12}>20,p_{T34}>15,m>20$ & 378 &195 &1.94\\
$p_{T12}>20,p_{T34}>15,m>10$ & 482 &260 &1.85\\
$p_{T12}>20,p_{T34}>10,m>10$ & 1418 &982& 1.44\\
$p_{T12}>20,p_{T34}>10,m>20$ & 966 &633 &1.53\\
$p_{T12}>15,p_{T34}>10,m>20$ & 1014 &673 &1.51\\
$p_{T12}>15,p_{T34}>10,m>10$ & 1497 &1048 & 1.43\\
$p_{T12}>10,p_{T34}>10,m>20$ & 1017 &675 &1.51\\
$p_{T12}>10,p_{T34}>10,m>10$ & 1503 &1056 &1.42\\
\hline
\end{tabular}
\newline
\newline
Table 3: Number of Events Passing Selection Criteria ($m<130$ GeV)
\newline
\newline
\begin{tabular}{ |p{5cm}|p{1cm}|p{1cm}|p{1cm}|p{1cm}|p{1cm}|
  }
\hline

\hline
Cuts& 0.3 &0.4 &0.5 &0.6 &0.7\\
\hline
$p_{T12}>20,p_{T34}>10,m>10$ & 625 &1418& 1220 & 869&606\\
$p_{T12}>15,p_{T34}>10,m>10$ & 666 &1497 &1285& 930&647 \\
$p_{T12}>10,p_{T34}>10,m>10$ & 667 &1503 &1291& 941&656\\
\hline
\end{tabular}
\newline
\newline
Table 4: Number of Events Passing Selection Criteria for $\Delta R = 0.3 - 0.7$
\newline
\linebreak
This lowered the selection efficiency but increased the signal-to-background ratio.  Using calculated values for the Higgs cross-section in an ep collider, a value for the background cross-section and the numbers of events computed in Tables 2 and 4, values were calculated for expected numbers of signal and background events along with the expected significance.  The actual signal-to-background ratios were much smaller as a result than the ones calculated initially, since the background cross-section is larger than the signal cross-section.
\newline
\newline
\begin{tabular}{ |p{6cm}|p{2cm}|p{3cm}|p{2cm}|
  }
\hline

\hline
Cuts& Signal &Background &Significance\\
\hline
$p_{T12}>20,p_{T34}>20,m>10$ & 568 &120400& 1.6\\
$p_{T12}>20,p_{T34}>20,m>20$ & 486 &95200 &1.6\\
$p_{T12}>20,p_{T34}>15,m>20$ & 1625 &546000 &2.2\\
$p_{T12}>20,p_{T34}>15,m>10$ & 2073 &728000 &2.4\\
$p_{T12}>20,p_{T34}>10,m>10$ & 6097 &2749600& 3.7\\
$p_{T12}>20,p_{T34}>10,m>20$ & 4154 &1772400 &3.1\\
$p_{T12}>15,p_{T34}>10,m>20$ & 4360 &1884400 &3.21\\
$p_{T12}>15,p_{T34}>10,m>10$ & 6437 &2934400 & 3.8\\
$p_{T12}>10,p_{T34}>10,m>20$ & 4373 &1890000 &3.2\\
$p_{T12}>10,p_{T34}>10,m>10$ & 6463 &2956800 &3.8\\
\hline
\end{tabular}
\newline
\newline
Table 5: Expected Numbers of Events ($m<130$ GeV)
\newline
\newline
\begin{tabular}{ |p{6cm}|p{2cm}|p{3cm}|p{2cm}|
  }
\hline

\hline
Cuts& Signal &Background &Significance\\
\hline
$p_{T12}>20,p_{T34}>20,m>10$ & 800 &249200& 1.6\\
$p_{T12}>20,p_{T34}>20,m>20$ & 692 &218400 &1.5\\
$p_{T12}>20,p_{T34}>15,m>20$ & 2030 &845600 &2.2\\
$p_{T12}>20,p_{T34}>15,m>10$ & 2554 &1094800 &2.4\\
$p_{T12}>20,p_{T34}>10,m>10$ & 7069 &3589600& 3.7\\
$p_{T12}>20,p_{T34}>10,m>20$ & 4902 &2366000 &3.2\\
$p_{T12}>15,p_{T34}>10,m>20$ & 5126 &2478000 &3.3\\
$p_{T12}>15,p_{T34}>10,m>10$ & 7443 &3788400 &3.8\\
$p_{T12}>10,p_{T34}>10,m>20$ & 5139 &2483600 &3.3\\
$p_{T12}>10,p_{T34}>10,m>10$ & 7473 &3816499 &3.8\\
\hline
\end{tabular}
\newline
\newline
Table 6: Expected Number of Events ($m<140$ GeV)
\newline
\newline
The background sample used at this point was for background due to W${-}$ jets and so the analysis was also run over a sample for more general QCD background.  We repeated the analysis for events with invariant 4-jet masses below 130 Gev and added in the effect of the second background.  The numbers of events were calculated and then scaled up appropriately.
\newline
\newline
\begin{tabular}{ |p{6cm}|p{2cm}|p{3cm}|p{2cm}|
  }
\hline

\hline
Cuts& Signal &Background &Significance\\
\hline
$p_{T12}>20,p_{T34}>20,m>10$ & 800 &453600& 1.2\\
$p_{T12}>20,p_{T34}>20,m>20$ & 692 &364400 &1.1\\
$p_{T12}>20,p_{T34}>15,m>20$ & 2030 &1838400 &1.5\\
$p_{T12}>20,p_{T34}>15,m>10$ & 2554 &2438000 &1.6\\
$p_{T12}>20,p_{T34}>10,m>10$ & 7069 &10218000& 2.2\\
$p_{T12}>20,p_{T34}>10,m>20$ & 4902 &6570800 &1.9\\
$p_{T12}>15,p_{T34}>10,m>20$ & 5126 &6945600 &1.9\\
$p_{T12}>15,p_{T34}>10,m>10$ & 7443 &11015400 &2.2\\
$p_{T12}>10,p_{T34}>10,m>20$ & 5139 &7038800 &1.9\\
$p_{T12}>10,p_{T34}>10,m>10$ & 7473 &11174899 &2.2\\
\hline
\end{tabular}
\newline
\newline
Table 7: Expected Number of Events with QCD Background ($m<130$ GeV)
\newline
\newline
\noindent
The number of background events is now quite high: choosing the signal for the first set of cuts in Table 6, the expected background events would have an uncertainty of $\sqrt{120400} = 347$ such that the signal would be seen at $1.5 \sigma$.  An improvement to $3 \sigma$ could likely be achieved with improvements to the analysis and use of a more refined selection strategy and improved technique for signal-to-background comparison.

\section{Conclusion}

\noindent
After cuts and analysis had been carried out, it was found that a good selection scheme for the search which we are studying is $|\Delta \eta| <1, 10<m<85$ GeV, $p_T$ of jets 1 and 2 between $10 - 20$ GeV and $p_T$ of jets 3 and 4 $>10$ GeV and that the selection efficiency is highest for $\Delta R = 0.4$, leading to an efficiency between $7.1 - 7.5 \%$ for finding the invariant 4-jet mass in a mass region $<140$ GeV.  In fact, $\Delta R = 0.4$ was optimal for all cut sets where it was varied, but this could be unique to the particular decay we were studying, as other Higgs decays often show a strong dependence on $\Delta R$.  It was also found that we could begin to incorporate a background sample due to W${-}$ jets with a signal which would be observed with $3.8 \sigma$, but attempts to improve this figure by adding QCD background resulted in a very high number of background events.

We should also point out this this is a preliminary study and that further work would be required before our conclusion could be stated more firmly.  The main reason is that we had not always been running on full signal statistics, only using 10000 events in the interest of efficiency when running samples with many modified versions of the analysis code to find effective cuts (something like a quarter of the full signal statistics).  We also did not account for the hadronic branching fraction of the W boson.  It is likely that both of these factors led to a small signal cross-section of around 0.1 pb.  In our calculations we have assumed a value of 1 ab of luminosity, whereas a revised value of 2 ab would double the signal cross-section.  This would however decrease the significance by a factor of $\sqrt{2}$ to compensate.  A proper comparison of signal-to-background would be difficult in this study and it might be considered satisfactory to have imposed a selection scheme and optimized $\Delta R$ but it would obviously be desirable to perform a more sophisticated signal-to-background analysis using boosted decision trees [4].  BDT is especially useful when the signal is 'drowned out' by similar-looking background events, since it can be used to identify if events are signal-like or background-like by using Monte Carlo simulations to train the decision tree.  This then enables a final determination of signal strength compared to background [6].  

It would also be desirable to adjust or refine cuts to increase the efficiency of signal detection, since the mass cuts employed appeared to be drowning out the signal at the region of interest.  The next step besides BDT would be to run the analysis again on the full Monte Carlo signal statistics, as this could have an effect on the signal cross-section.  The effect of other variables could be studied for both signal and background: in particular, adjustments of $\Delta \eta$ were not investigated for background topologies in this study.  The study only considered charged current and QCD background.  In a more thorough study, different types of neutral current background could be incorporated.

\section*{Acknowledgements} 
\noindent
I would like to thank my supervisors Mario Campanelli and Uta Klein for useful discussions and for providing data samples.

\section*{References}

[1] J. Alwall  et al. The automated computation of tree-level and next-to-leading order differential cross sections, and their matching to parton shower simulations. JHEP, 1405.0301:185:2250–2300, 2014.

\noindent [2] J. Esteves.  Understanding the Higgs boson with the Large Hadron Electron Collider. \textit{J. Phys}.: Conf. Ser, 645:012009, 2015.

\noindent [3] A. Giammanco et al. DELPHES 3, A modular framework for fast simulation of a generic collider experiment. JHEP, 02:057:1307.6346, 2014.

\noindent [4] Hoecker et  al.  TMVA:  Toolkit  for  Multivariate  Data  Analysis. PoS, 

\noindent ACAT:040:physics/0703039, 2007.

\noindent [5] LHeC Study Group.   A Large Hadron Electron Collider at CERN: Report on the Physics and Design Concepts for Machine and Detector. 2012.

\noindent [6] G. McGregor et al. Boosted decision trees:  an alternative to artificial neural networks. Nucl. Instrum. Meth, A543(2-3):577-584:physics/0408124, 2005.

\noindent [7] S.  Mrenna,  T  Sjostrand  and  P.  Skands.   PYTHIA  6.4  Physics  and  Manual. JHEP, 05:026,:hep.ph/0603175, 2006.

\noindent [8] C. Patrignani  et al. Review of Particle Physics. \textit{Chin. Phys. C}, 

\noindent 40:100001,2016/2017.

\end{document}